\documentstyle[twoside,12pt,epsf]{article}
\normalsize
\newcommand{\bdi}{\begin{displaymath}}
\newcommand{\edi}{\end{displaymath}}
\newcommand{\bfi}{\begin{figure}}
\newcommand{\efi}{\end{figure}}

\newcommand{\beq}{\begin{equation}}
\newcommand{\eeq}{\end{equation}}
\newcommand{\beqa}{\begin{eqnarray}}
\newcommand{\eeqa}{\end{eqnarray}}

\newcommand{\ra}{\rightarrow}

\def\longbar#1{\setbox1=\hbox{$#1$}
\setbox2=\vbox{\hrule width 0.8\wd1}
\raise0.5\ht1\hbox{${\lower\dp1\box2}\atop\box1$}}  

\begin{document}

\begin{titlepage}

\begin{flushright}
\today
\end{flushright}

\vspace{1cm}
\begin{center}
{\Large \bf Particle creation via relaxing hypermagnetic knots}\\[1cm]
C. Adam* \\
Institut f. theoretische Physik, Universit\"at Karlsruhe, 76128 Karlsruhe \\

\medskip

\medskip

B. Muratori**,\, C. Nash*** \\
Department of Mathematical Physics, National University of Ireland, Maynooth
\vfill
{\bf Abstract} \\
\end{center}

We demonstrate that particle production for fermions coupled chirally to
an Abelian gauge field like the hypercharge field is provided by the
microscopic mechanism of level crossing. For this
purpose we use recent results on zero modes of Dirac operators 
for a class of localized hypermagnetic knots.

\vfill

$^*)${\footnotesize  
email address: adam@maths.tcd.ie, adam@pap.univie.ac.at} 

$^{**})${\footnotesize
email address: bmurator@fermi1.thphys.may.ie} 

$^{***})${\footnotesize
email address: cnash@stokes2.thphys.may.ie} 
\end{titlepage}

\section{Introduction}

It is well-known that in chiral theories 
the chiral anomaly \cite{Ad1,BeJa}  together
with some topological properties of the gauge fields open the
possibility for particle-number violating processes \cite{tH1}. 
On one hand, such
processes may become relevant at high-energy experiments in the TeV
range, on the other hand some version of these particle-number violating
processes is believed to be the most promising candidate for explaining
the dominance of matter over antimatter observed in the current
universe.

These particle-number violating processes are described in a slightly
different manner, depending on whether the relevant gauge fields are
non-Abelian (containing an SU(2) subgroup) or Abelian. In the
non-Abelian case even gauge fields interpolating between vacuum
configurations at initial time $t_{\rm i}$ and final time $t_{\rm f}$
may provide particle-number violation (i.e. particle creation) provided
that the pure vacuum fields at $t_{\rm i}$ and $t_{\rm f}$ have
different winding numbers $n_{\rm i}$ and $n_{\rm f}$,
respectively. Moreover, the total number of particles created is two
times the Pontryagin index of the given gauge field in a Euclidean
formulation. In Minkowski space this translates into the condition that
the number of created particles is two times the number of levels
(eigenvalues of the spatial but time-dependent Dirac operator) that
cross zero as the gauge field evolves from the vacuum at $t_{\rm i}$   
to the vacuum at $t_{\rm f}$ \cite{Chr1}. 
This introduces a microscopic description
of the particle creation mechanism in the sense that each time a level
crosses zero a particle is created and an anti-particle is annihilated
(or vice versa, depending on the sign of the level crossing).

In the Abelian case there are no nontrivial, nonequivalent vacuum
configurations, and therefore particle creation is not expected for
vacuum to vacuum transitions. However, the anomaly relation still
predicts particle creation, e.g., for a transition from a non-vacuum
configuration (with non-zero Chern--Simons number) to a vacuum
configuration \cite{Vil1,RuTa1}. 
It may for instance happen that  non-trivial configurations of the
Abelian hyper-charge field (hypermagnetic knots) are formed at an
instance of time when there is still thermodynamic equilibrium in the
universe such that any imbalance between particles and antiparticles is
washed out immediately. When these hypermagnetic knots relax at later
times when thermodynamic equilibrium no longer holds, particle creation
may result. These features are described in detail e.g. in 
\cite{JoSh1}--\cite{Giov2} and in the references quoted there.
 
Here the question arises whether a similar microscopic
mechanism (like the level-crossing phenomenon) may be identified that is
responsible for the particle creation. The problem here is that there is
no simple topological feature like the index theorem 
\cite{AS1,JR1,JR2} in the
non-Abelian case that guarantees the existence and counts the number of
zero modes. In addition, until recently there was not much information
available on zero modes of the spatial (i.e. three-dimensional) Abelian
Dirac operator, at all. The first example of such a zero mode was given
only in 1986 in \cite{LoYa}, and some further results including the
feature of zero mode degeneracy were obtained recently in
\cite{ABN1}--\cite{BaEv}.

It is the purpose of this paper to show that in the case of the class
of localized hypermagnetic knots that were discussed in \cite{ABN3} 
it is indeed the level crossing
phenomenon that accounts for particle production. We shall find that
the number of levels crossed exacly matches the number of particles
created as predicted by the anomaly equation.

We use Minkowski space conventions, $\eta_{\mu\nu} = {\rm diag}\,
(1,-1,-1,-1)$,  Greek indices are space-time indices
and latin indices are space indices.

\section{Anomaly equation}

The Abelian anomaly equation for one fermion species is (see e.g. 
\cite{Bertl})
\beq
\partial^\mu (J^{\rm L}_\mu - J^{\rm R}_\mu )=\frac{1}{16 \pi^2}
\epsilon^{\mu\nu\alpha\beta}F_{\mu\nu}F_{\alpha\beta}
\eeq
where the coupling constant is absorbed into the Abelian gauge field
$A_\mu$, $F_{\mu\nu}=\partial_\mu A_\nu - \partial_\nu A_\mu$ and it is
assumed that the left-handed and right-handed currents couple
differently. Here one could, for example, 
assume the hyper-charge assignments of
the different particle species. However, as we are mainly interested in 
the general features we
assume in the sequel that the coupling is purely chiral, $J^{\rm R} =0$,
and we consider only one particle species.
Further we assume for later convenience that $A_0 =0$. Then the anomaly
equation may be re-written like
\beq
\partial^\mu J^{\rm L}_\mu =-\frac{1}{4\pi^2}\epsilon_{ijk} [\partial_0
(A_i \partial_j A_k) + \partial_i (A_j \partial_0 A_k)] 
\eeq
(concerning the apparent sign change, we use Euclidean conventions for
the purely space-like indices, i.e. $a_j b_j \equiv \vec a\cdot \vec b$).
If we further assume that both currents and gauge fields are confined to
a finite space region for the time interval we consider we find upon
integration over all space and the time interval $[t_{\rm i},t_{\rm f}]$
\beq
\int d^3 x (J^{\rm L}_0 (t_{\rm f},\vec x) - J^{\rm L}_0 (t_{\rm i},\vec x))
=-\frac{1}{4\pi^2}\int d^3 x [(\vec A\cdot \vec B)(t_{\rm f},\vec x)
-(\vec A\cdot \vec B)(t_{\rm i},\vec x)]
\eeq
where the l.h.s. is the change in particle number between $t_{\rm i}$
and $t_{\rm f}$. 

\section{Level crossing: simplest case}

For the discussion of the level crossing phenomenon, let us start from
the usual four-dimensional Dirac equation
\beq
\gamma^\mu (-i\partial_\mu -A_\mu)\psi =0 .
\eeq
We want to study zero energy bound states at some fixed time, therefore
$\psi$ has to be independent of time for our purposes. Further, we study
a chiral theory, therefore the four-component Dirac spinor $\psi$ has to
be replaced by the two-component Weyl spinor $\Psi$. Assuming $A_0 =0$
and choosing an approriate representation of the gamma matrices we
finally get ($\vec \sigma$ are the Pauli matrices)
\beq
\sigma_k (-i\partial_k -A_k )\Psi =0.
\eeq
Here $A_k$ consists of space components only but is still time
dependent. As we want to study relaxation processes that do not change
the shape of the gauge field (and hypermagnetic field) we now assume
\beq
A_k (t,\vec x) =c(t)A_k (\vec x)
\eeq
where $c(t)$ relaxes from an initial to a final value.

At this point we should describe the simplest knotted field
configuration that we want to discuss in this section. The gauge field and its
hypermagnetic field are ($r:=|\vec x|$)
\beq
A_l (t,\vec x) =c(t) \frac{N_l (\vec x)}{1+r^2}
\eeq
\beq
B_j (t,\vec x) =\epsilon_{jkl}\partial_k A_l (t,\vec x) =
4 c(t)\frac{N_j (\vec x)}{(1+r^2)^2} 
\eeq
where $\vec N (\vec x)$ is the vector field 
\beq
\vec N (\vec x) = \frac{1}{(1+r^2)} 
\left( \begin{array}{c} 2x_1 x_3 -2x_2  \\ 2x_2 x_3 +2x_1 \\
1-x_1^2 -x^2_2 +x_3^2 \end{array} \right) 
\eeq
with unit length $\vec N^2 =1$.
In addition, there is a hyperelectric field due to the time dependence of
$A_k$, which may however be weak if $c(t)$ changes only slowly in time.

The hypermagnetic field  (8) is the simplest possible localized knotted
magnetic field configuration \cite{Giov2}. 
Alternatively it may be described as the
Hopf curvature of the simplest Hopf map \cite{ABN3,Ran1}, 
or it may be generated
in the following way: choose the simplest nontrivial SU(2) pure
gauge element $U(\vec x)$ on compatified ${\rm R}^3$ with winding number
one (i.e. the identity map $S^3 \ra S^3$). 
Perform an Abelian projection via $G_k (\vec x)={\rm tr} \, (U^\dagger
\partial_k U \sigma_3 )$, then the Abelian gauge field $G_k$ is proportional
to the gauge field (7), and the magnetic field $\epsilon_{jkl}\partial_k
G_l$ is proportional to the knotted hypermagnetic field (8),
\cite{JaPi}.

Now we need the following important result on zero modes of the above
spatial Dirac equation (5) for the specific gauge field (7). When the
factor $c(t)$ in (7) (treated now as an arbitrary constant) obeys
\beq
c = 1+2k \, , \quad k=1,2,3,\ldots
\eeq
then the Dirac equation (5) has precisely $k$ (square-integrable,
non-singular) zero modes. This was demonstrated by explicit construction
of the zero modes in \cite{ABN2,ABN3}. Further, it was proven in
\cite{ErSo} that these are indeed all zero modes that exist for the
Dirac operator (5) with gauge field (7). 

For later convenience, we display the single zero mode for the simplest
case, $c=3$,
\beq
\Psi =\frac{4}{(1+r^2)^{\frac{3}{2}}}({\bf 1} +i\vec x \vec \sigma )
\left( \begin{array}{c} 1  \\ 0 \end{array} \right) 
\eeq
The construction of the higher zero modes (for $c=5,7,\ldots $) is
explained in the appendix.

Now assume that $c(t)$ is relaxed from an initial value
\beq
c(t_{\rm i})=1+2K
\eeq
where $K$ is assumed large, to a small (or zero) final value $c(t_{\rm
  f}) < 3$ such that $c' <0$ for $t_{\rm i}\le t \le t_{\rm f}$. Then
the number $L$ of levels crossed obviously is
\beq
L=\sum_{k=1}^K k=\frac{K(K+1)}{2} .
\eeq
Finally we have to calculate the Chern--Simons number of the gauge field
(7). This is easily done and leads to
\beq
\frac{1}{4\pi^2}\int d^3 x \vec A(t,\vec x)\cdot \vec B(t, \vec x) 
=\frac{1}{4\pi^2}\int d^3 x \frac{4c^2 (t)}{(1+r^2)^3}
= \frac{c^2 (t)}{4} .
\eeq
Inserting this into the anomaly equation (3) and using the value (12) for
$c (t_{\rm i})$ leads to the following result for the number $N$ of
particles created
\beq
N=\frac{1}{4}(c^2 (t_{\rm i}) - c^2 (t_{\rm f}))=K(K+1) +\frac{1}{4}(1-c^2
(t_{\rm f}))
\eeq
Interestingly, if we want to have a precise matching $N=2L$ between the
number $N$ of created particles and the number $L$ of level crossings
then we should choose $c(t_{\rm f}) =\pm 1$ instead of $c(t_{\rm f}) =0$
(in fact, $c(t_{\rm f})=-1$ will turn out to be the more natural choice,
see below).
However, for large $K$ this difference becomes, of course, negligible.

\section{Level crossing: more general cases}

Here we want to demonstrate that the relation $N=2L$ between the number
$N$ of particles created and the number $L$ of levels crossed holds, in
fact, for a much wider class of gauge potentials (and their
hypermagnetic fields). This class of fields and their zero modes were
discussed in \cite{ABN3}, so we have to review some of these results. 

In \cite{ABN3} the concept of Hopf maps was used, so let us briefly
explain it.
 Hopf maps are maps $S^3 \ra S^2$. 
The third homotopy group of the two-sphere is non-trivial,
$\Pi_3 (S^2) ={\bf Z}$, therefore such maps are characterised by an
integer topological index, the so-called Hopf index.
Hopf maps may be expressed, e.g., by maps $\chi :
{\rm\bf R}^3 \ra {\rm\bf C}$ provided that the complex function $\chi$
obeys $\lim_{|\vec x|\ra \infty} \chi (\vec x)=\chi_0 ={\rm const}$.
The pre-images in ${\rm\bf R}^3$
of points of the target $S^2$ (i.e., the pre-images of points $\chi =
{\rm const}$) are closed curves in ${\rm\bf R}^3$ (or in the 
related domain $S^3$). Any two different closed curves 
are linked $N_{\rm H}$ times,
where $N_{\rm H}$ is the Hopf index of the given Hopf map $\chi$.
Further, a magnetic field $\vec {\cal B}$ 
(the Hopf curvature) is related to the
Hopf map $\chi$ via
\beq
\vec {\cal B} = 
\frac{2}{i}\frac{(\vec\partial\bar\chi)\times(\vec\partial\chi)}{
(1+\bar\chi \chi)^2} =2\frac{(\vec\partial T)\times\vec
\partial\sigma}{(1+T)^2}
\eeq
where $\chi =Se^{i\sigma}$ is expressed in terms of its modulus $S=:T^{1/2}$
and phase $\sigma$ at the r.h.s. of (16).

Mathematically, the curvature ${\cal F}=\frac{1}{2}{\cal F}_{ij}dx_i dx_j$, 
${\cal F}_{ij}= \epsilon_{ijk}{\cal B}_k$, 
is the pullback under the Hopf map, ${\cal F}=\chi^* \Omega$,
of the standard area two-form $\Omega$ on the target $S^2$. 
Geometrically, $\vec {\cal B}$ is 
tangent to the closed curves $\chi ={\rm const}$ (see e.g. 
\cite{Ran1,FN1,BS1,JaPi}).
The Hopf index $N_{\rm H}$ of $\chi$ may be computed from $\vec {\cal B}$ via
\beq
N_{\rm H}=\frac{1}{16\pi^2}\int d^3 x \vec {\cal A} \cdot \vec {\cal B}
\eeq
where $\vec {\cal B}=\vec\partial\times \vec {\cal A}$.

The simplest (standard) Hopf map $\chi$ with Hopf index $N_{\rm H}=1$ is
\beq
\chi =\frac{2(x_1 +ix_2)}{2x_3 -i(1-r^2)}
\eeq
with modulus and phase 
\beq
T:= \bar\chi \chi = \frac{4(r^2 -x_3^2)}{4x_3^2 +(1-r^2)^2}\, ,\quad
 \sigma = \arctan
\frac{x_2}{x_1} + \arctan\frac{1-r^2}{2x_3} .
\eeq
The Hopf curvature of the simplest standard Hopf map is 
\beq
\vec {\cal B} = \frac{16}{(1+r^2)^2} \vec N
\eeq
i.e., it is just the magnetic field (8) with $c=4$.

Specifically, we are interested in Hopf maps that are compositions of
the standard Hopf map with maps $G: S^2 \ra S^2$,
\beq
\chi_G :S^3 \stackrel{\chi}{\ra} S^2 \stackrel{G}{\ra}S^2 .
\eeq
Here, if $G$ has
degree (i.e. winding number) $m$, then the composed Hopf map $\chi_G$
has Hopf index $N_{\rm H}=m^2$.

Now the result of \cite{ABN3} is as follows. Construct a gauge field
(and its magnetic field) by subtracting the ``background'' field $\vec
A^{\rm B}$ from an arbitrary Hopf connection of the type (21),
\beq
\vec A^{(G)} = \vec {\cal A}^{(G)} - \vec A^{\rm B}
\eeq
\beq
\vec B^{(G)} = \vec \partial \times \vec A^{(G)} =
\vec {\cal B}^{(G)} - \vec B^{\rm B}
\eeq
where $\vec {\cal B}^{(G)}$ is the Hopf curvature of the Hopf map (21) and
\beq
\vec A^{\rm B} =\frac{1}{1+r^2}\vec N
\eeq
is just the simplest gauge field (7) (with $c=1$).
Then the Dirac equation (5) with gauge field $\vec A^{(G)}$ has
\beq
k=2m-1
\eeq
zero modes. 
This implies that half-integer $m$ corresponding to 
double-valued, square-root type maps $G: S^2 \ra S^2$ have to be allowed
in order to take into account the cases when the number of zero modes is
even.

Again, the zero modes for gauge fields of the type (22)
have been constructed explicitly in \cite{ABN3},
and it was proven in \cite{ErSo} that these are all zero modes that
exist for the given gauge fields.

As in \cite{ABN3} we now want to consider in detail the class of maps
$G$
\beq
G(z,\bar z)=f(\bar z z)z^m =: g^{1/2}(\bar z z)e^{im \arg (z)}
\eeq
(we use the complex variable $z$ as a
stereographic coordinate on $S^2$) which is
indeed a map $S^2 \ra S^2$ with winding number $m$ 
provided that $g(0)=0$ and $g(\infty )=\infty
$, see \cite{ABN3}. Concretely, we assume
\beq
\lim_{|z|\to 0}g(\bar z z) \sim |\bar z z|^{c_0} \, , \quad c_0 >0
\eeq
\beq
\lim_{|z|\to \infty }g(\bar z z) \sim |\bar z z|^{c_\infty} \, , 
\quad c_\infty >0
\eeq 
Otherwise $g$ is not much restricted (of course, $g\ge 0$ holds by
definition). The corresponding Hopf map with Hopf index $N_{\rm H}=m^2$
reads
\beq
\chi_{g,m}=g^{1/2}(T)e^{im\sigma}
\eeq
where $m$ is integer or half-integer, and $T$ and $\sigma$ are given in
(19). The Hopf curvature of the Hopf map (29) is ($'\equiv (\partial
/\partial T)$)
\beq
\vec {\cal B}^{(g,m)}=m\frac{g'(1+T)^2}{(1+g)^2}\vec {\cal B}=:
m \vec {\cal B}^{(g)}
\eeq
where $\vec {\cal B}$ is the simplest Hopf curvature (20). The Hopf
connection  of (30) is
\beq
\vec {\cal A}^{(g,m)} = m \vec {\cal A}^{(g)}= m\Bigl(
\frac{4}{1+r^2}\vec N +\frac{1}{T}(\frac{1}{1+g}-\frac{1}{1+T})(\vec
\partial T) \times \vec N\Bigr) .
\eeq
It may be checked after some algebra that indeed $\vec {\cal B}^{(g,m)}
= \vec \partial \times \vec {\cal A}^{(g,m)}$, but we shall compute $
\vec {\cal A}^{(g)}$ in a simple fashion in the appendix. In addition,
the construction of the zero modes for the gauge fields (22) with Hopf
maps (29) is demonstrated explicitly in the appendix.

Now let us construct the following time-dependent gauge fields,
\beq
\vec A^{(G)}(t) = c(t) \vec {\cal A}^{(g)} - \vec A^{\rm B}
\eeq
\beq
\vec B^{(G)}(t) = c(t) \vec {\cal B}^{(g)} - \vec B^{\rm B}
\eeq
where $c(t)$ relaxes from a large initial value $c(t_{\rm i})$ to zero,
$c (t_{\rm f}) =0$. If
\beq
c(t_{\rm i})=\frac{K+1}{2}
\eeq
then the total number of levels crossed between $t_{\rm i}$ and $t_{\rm
  f}$ is again
\beq
L=\sum_{k=1}^K k =\frac{K(K+1)}{2} .
\eeq
Now we have to evaluate the Chern--Simons integral in (3),
\bdi
N=\frac{1}{4\pi^2}\int d^3 x \vec A^{(G)}(t_{\rm i}) \cdot
\vec B^{(G)}(t_{\rm i}) - 
\frac{1}{4\pi^2}\int d^3 x \vec A^{(G)}(t_{\rm f}) \cdot
\vec B^{(G)}(t_{\rm f})
\edi
\bdi
=\frac{c^2 (t_{\rm i})}{4\pi^2} \int d^3 x \vec {\cal A}^{(g)}\cdot
\vec {\cal B}^{(g)} - \frac{c(t_{\rm i})}{4\pi^2} \int d^3 x
[ \vec {\cal A}^{(g)}\cdot \vec B^{\rm B}+
 \vec A^{\rm B} \cdot \vec {\cal B}^{(g)} ]
\edi
\beq
=\frac{c^2 (t_{\rm i})}{4\pi^2} \int d^3 x \frac{64}{(1+r^2)^3}
\frac{g' (1+T)^2}{(1+g)^2} -
\frac{c(t_{\rm i})}{4\pi^2} \int d^3 x [\frac{16}{(1+r^2)^3}
 +\frac{16}{(1+r^2)^3} \frac{g' (1+T)^2}{(1+g)^2} ] .
\eeq
Here we may use the fact that $\vec {\cal B}^{(g)}$ is a Hopf curvature
with Hopf index one,
\beq
\frac{1}{16\pi^2} \int d^3 x \vec {\cal A}^{(g)}\cdot
\vec {\cal B}^{(g)} =1
\eeq
and the fact that the two integrands in (36) that are multiplied by 
$c(t_{\rm i})$ are proportional to the integrand in
(37) and to the simplest Chern--Simons integrand in (14), respectively. We
arrive at
\beq
N=4c^2 (t_{\rm i}) - c (t_{\rm i}) (1+1) = 4 (\frac{K+1}{2})^2
-2\frac{K+1}{2} = K(K+1) .
\eeq
Therefore, the relation $N=2L$ between the number $N$ of particles
created and the number $L$ of levels crossed is confirmed for the
class of gauge fields (32) with the knotted hypermagnetic fields (33).

Observe that only the Hopf connection part in (32) is relaxed from
$c(t_{\rm i})$ to zero, whereas the ``background'' field remains
unchanged. However, for sufficiently large $c(t_{\rm i})$ the difference
becomes negligible. In addition, as the Hopf curvatures of all the Hopf
maps (21) point into the same direction as the background field
$\vec B^{\rm B}$ at each point in space, the hypermagnetic fields (23) (and
specifically (33)) are
still knotted field configurations.  

\section{Summary}

For a whole class of localized hypermagnetic knots we have
demonstrated that the particle creation that is predicted by the anomaly
equation is indeed provided by the microscopical mechanism of level
crossing. We found a precise matching between the number of
levels crossed on one hand and the number of particles created on the
other hand. This precise matching leads, of course, to the immediate
conjecture that this microscopic description remains true in more
general cases. However, as already stated, some
major results on zero modes of the
Abelian Dirac operator in three dimensions were obtained only recently,
therefore more mathematical investigations are necessary before this question
can be finally answered. 

\section*{Acknowledgments}
BM gratefully acknowledges financial support from the Training and 
Mobility of Researchers scheme (TMR no. ERBFMBICT983476).

\section*{Appendix}

In this appendix we want to construct the zero modes of the Dirac
operators (5) with gauge fields (22) for the specific class of Hopf maps
(29). For this purpose we shall make use of some further results of
\cite{ABN3}, so let us briefly review them. It was shown in \cite{ABN3}
that the spinor
\beq
\Psi^{(M)} = e^{M/2}\Psi
\eeq
is a zero mode for some Dirac operator. Here $\Psi$ is the simplest zero
mode (11) and $M$ is a function of $T$ only, $M=M(T)$ ($T$ is given in
(19)). Further it should hold that $\exp (M)>0$ for all $T<\infty$, i.e.,
$\exp (M)$ has no zeros at finite $T$. If $\exp (M)$ behaves like
\beq
\lim_{T \to \infty}\exp (M) \sim T^{-n_\infty}
\eeq
(where $n_\infty$ must be a non-negative integer) then further
square-integrable zero modes for the {\em same} Dirac operator like
$\Psi^{(M)}$ are
\beq
\Psi^{(M)}_l =\chi^l \Psi^{(M)} \, , \quad l = 0, \ldots , n_\infty
\eeq
and there exist $k=n_\infty +1$ zero modes for this Dirac
operator. The task in \cite{ABN3} was to calculate the Dirac operator
and the related magnetic field from a given zero mode $\Psi^{(M)}$, and
to show that these magnetic fields are indeed related to Hopf curvatures
as indicated in (23). 

Here we want to do the opposite. We assume that a Hopf curvature $\vec
{\cal B}^{(g,m)}$ (i.e., a function $g$ and a integer or half-integer
number $m$) with Hopf index $N_{\rm H}=m^2$ is given, and we want to
find $M$ and the number of zero modes. In fact, the relation between
$(g,m)$ and $(M,n_\infty)$ was already calculated in \cite{ABN3} and
reads
\beq
\frac{m}{1+g} =\frac{1}{1+t} + \frac{1}{2}TM' +\frac{1}{2}n_\infty
\eeq
\beq
m=1+\frac{1}{2}n_\infty
\eeq which may be re-expressed as
\beq
M' = \frac{1}{T} [\frac{n_\infty +(2+n_\infty )T -2g}{(1+g)(1+T)}]
-\frac{n_\infty}{T} .
\eeq
If $g$ behaves as is assumed in (27,28) then (44) is integrable at
$T=0$. Further, it is integrable for finite $T$ (even if $g$ has a
singularity or zero somewhere), which guarantees $-\infty <M<\infty $ for
$T<\infty $, as is assumed above. In addition,
the large $T$ behaviour is determined entirely by the
second term on the r.h.s. of (44) and leads to $M\sim -n_\infty \ln T$ as
it must be. 

Now we want to use (44)  to prove expression (31) for the Hopf
connection $\vec {\cal A}^{(g)}$. It was shown in \cite{ABN3} that 
$\vec {\cal A}^{(g)}$ in terms of $M$ reads
\beq
\vec {\cal A}^{(g)} =\frac{4}{1+r^2}\vec N +\frac{1}{2}M' (\vec
\partial T) \times \vec N .
\eeq
$\vec {\cal A}^{(g)}$ is a Hopf connection with Hopf index
one, therefore we may insert $M'$ from (44) with $n_\infty =0$ into (45),
which easily reproduces (31)
(a further pure gauge contribution is absent in (45) because due to
$n_\infty =0$ it holds that $\exp (M) >0
\, \forall \, T$).

Finally, we may calculate the zero modes for the fields of Section 3
(the simplest hypermagnetic knots) as an example. Here $g=T$ and
\beq
M= - n_\infty \ln (1+T)
\eeq
(where we chose the irrelevant integration constant such that $M(0)=0$).
Therefore, if the constant $c$ in (10) is $c=1+2k$, the $k=n_\infty +1$
zero modes are
\beq
\Psi^{(M)}_l = \chi^l (1+T)^{-\frac{n_\infty}{ 2}}\Psi \, , \quad l=0, \ldots
,n_\infty 
\eeq
where $\Psi$ is given in (11).

\end{document}